\documentstyle[aps,prb]{revtex}
\begin{document}
\title{THEORY OF OPTICAL SPECTRA OF POLAR QUANTUM WELLS: TEMPERATURE EFFECTS}
\author{F.J.Rodr\'{\i}guez*}
\address{Departamento de F\'{\i}sica, Universidad de los Andes, AA 4976, 
Bogot\'a D.C., Colombia\\
}
\maketitle
\begin{abstract}
Theoretical and numerical calculations of the optical absorption spectra of excitons interacting with 
longitudinal-optical phonons in quasi-2D polar semiconductors are presented. 
In II-VI semiconductor quantum wells, exciton binding energy can be 
tuned on- and off-resonance with 
the longitudinal-optical phonon energy by varying the quantum well width.
A comprehensive picture of this tuning effect on the
temperature-dependent exciton absorption spectrum is derived, using the
exciton Green's function formalism at finite temperature. The
effective exciton-phonon interaction is included in the
Bethe-Salpeter equation. 
 Numerical results are illustrated for $ZnSe$-based quantum wells. 
At low temperatures, both a single exciton peak as well as a continuum resonance state are found in 
the optical absorption spectra. By constrast,
 at high enough temperatures, a splitting of the exciton line due to 
real phonon absorption processes is predicted.
Possible previous experimental observations of this splitting are discussed. 

\end{abstract}

\vspace{0.5cm}
\section{Introduction}

Low-dimensional semiconductor structures provide with many surprising 
properties that have never been observed in bulk materials. One of the most 
interesting properties of II-VI semiconductor based nanostructures is the 
observation of well-resolved heavy- and light-hole exciton peaks 
in the absorption spectra, 
even at room temperature.\cite{ding} It has been proposed that the
lasing mechanism in highly excited II-VI quantum wells (QWs) corresponds to a coherent
recombination of excitons.\cite{ding,cingo1,pellegrini,ehrenreich}. It is therefore expected that 
excitons remain stable in the presence of high carrier  densities,
thus playing an important role in the operation of blue-green diode lasers. However, 
the potentialities
of II-VI compound heterostructures to obtain reliable excitonic devices 
depend strongly upon the strength, the width of the resonances  
and the nature of the exciton scattering processes at room temperature.
Indeed one of the main scattering mechanisms to be considered
is the interaction of excitons with longitudinal-optical (LO) phonons,\cite{poweleit} a problem of 
an enormous interest from both the experimental and theoretical point of views. 

A major difference between III-V and II-VI semiconductors is the highly polar
character of the latter ones. As a consequence, the coupling of charge carriers
(electrons and holes) with LO-phonons is stronger in II-VI than
in III-V compounds. A convenient measure of the coupling strength is the dimensionless
Fr\"ohlich constant $\alpha$. The electron-phonon complex represents a quasiparticle, called a polaron, which
in the weak coupling limit, $\alpha \ll 1$ as appropriate for $GaAs$ ($\alpha = 0.06$),
slightly renormalizes the electron properties.  II-VI compounds like $ZnSe$
should be described by an intermediate coupling theory since for this last
compound $\alpha = 0.43$, about an order of magnitude larger than in $GaAs$.  
Besides that, the exciton binding energy, $E_X$, in II-VI QWs can be made comparable to or even greater
than the LO-phonon energy, $\omega_{LO}$ (here and in the following $\hbar=1$), by varying the well width. 
In particular, for $ZnSe$, the binding energy of the bulk exciton $E_X^{3D}\approx 20 meV$  
while in two dimensions $E_X^{2D} = 4 E_X^{3D}$; the LO-phonon energy is $\omega_{LO}\approx 30 meV$ in bulk.
Therefore, three regimes can be expected in II-VI low dimensional structures: (1) $E_X > \omega_{LO}$; 
(2) $E_X < \omega_{LO}$ and (3) $E_X \approx \omega_{LO}$, to which we will refer from now on as {\it the
resonant regime}. 
Cases (1) and (2) have been extensivelly considered: for the first case,  a single LO phonon absorption cannot
dissociate the exciton into the electron-hole continuum. Furthermore, the exciton
Bohr radius is smaller than the polaron radius, consequently the electron-hole pair is weakly influenced by
phonons, i.e. they interact through a Coulomb interaction screened by  the static dielectric 
constant.\cite{varmasak} 
Hence, the coupling of excitons with LO-phonons via the Fr\"ohlich interaction, can be substantially reduced 
yielding to strong exciton absorption, even at 
room temperature\cite{ding,cingo1,pellegrini,ehrenreich,pelekanos,pelekanos2,oneil}. 
For the second case, the electron-hole Coulomb interaction 
will be screened by the lattice vibrations and other excitations\cite{varmasak,koinov}. Therefore, the exciton 
should be easily ionized affecting the optoelectronic properties of the semiconductor system.

The third case at finite temperature, to the best of our knowledge, has been ignored in theoretical studies.
 In the limit of vanishing temperature, previous works on II-VI heterostructures show that the 
exciton-LO phonon interaction weakens
with decreasing well width and predict a small feature in the exciton continuum spectrum
the so-called exciton-LO phonon quasi-bound state.\cite{pelekanos,pelekanos1}  
It is explained based on the large Fr\"ohlich coupling between excitons and LO-phonons in those compounds.
Similar but weaker effects have also been reported
for II-VI bulk systems\cite{toyozawa,hermanson,trallero}.  
Clearly, there is a lack of theoretical works on exciton-LO-phonon complexes in the resonant regime at 
high enough temperatures ($77 K \leq T \leq 330 K$) for such materials.

 The main goal of this work is to study the exciton absorption lineshape in the resonant regime 
as a function of temperature. 
 {\it At high temperatures} a peak splitting is predicted, 
corresponding to exciton-LO-phonon absorption processes. In the resonant regime both of these 
splitted peaks have roughly the same oscillator strength.  Therefore, this new feature should be 
considered for understanding the structure and dynamics of the optical response of II-VI QWs.\cite{umlauff}
 Broadening and splitting of the exciton peak at the lower band edge in III-V bulk 
systems have been reported,\cite{donlagic} and they have been explained as a consequence of the 
increasing of the phonon density of states. However, the exciton splitting
in this last system is vaguely visible for two reasons: {\it (i)} the Fr\"ohlich coupling is roughly one order of
magnitude lower than in II-VI heteroestructures; {\it (ii)} the resonance between exciton binding energy and
LO-phonon energy can not be achieved. By contrast, in our systems these last two points are met, yielding to
an enhancement of the oscillator strength. Therefore, the exciton splitting should become clearly visible in
II-VI heterostructures.

In the present work a theory of exciton-LO-phonon complexes using a 
Green's function (GF) formalism at finite temperature is developed. Using this formalism the linear optical absorption 
for II-VI-based QWs is obtained.\cite{ssr,mahan} Most of previous theoretical works employ the 
variational formalism to obtain exciton properties\cite{castella,zheng,gerlach,chuu},
with severe limitations to fit the whole absorption spectrum. 
A crucial difference between the present work and previous 
studies is that
the GF method employed here allows to include all the information about
both bound and continuum exciton states, going clearly
beyond the usual variational treatments. As a consequence, the whole
absorption spectrum, instead of just ground state properties, can be determined.
 Although there is a few
number of theoretical calculations on the linear optical absorption using the GF 
formalism\cite{koinov,pelekanos1,ssr}, they do not explore the joint effects: temperature and
QW width variations. 
The GF formalism used here allows a comprehensive picture of light absorption, 
treating quasiparticles and their interactions on the same footing.

The article is organized as follows: Sec. II gives a brief survey of GF formalism
 as applied  to the exciton-LO-phonon problem; in particular self-energy and vertex contributions as well as
the effective nonlocal dynamical electron-hole
 potential are described. In Sec. III the numerically calculated absorption
spectra for ZnSe/ZnSeS QWs are discussed. In Sec. IV  
vertex and self-energy contributions to the optical absorption are separately considered . 
Finally, Sec. V summarizes the main conclusions of this work.

\section{Theoretical Model}

Using a Fr\"ohlich Hamiltonian\cite{frolich} an 
electron-hole pair confined in an infinite barrier potential QW of width
$a$ ($z$ direction) is considered. In this work only the lowest electron and heavy-hole subbands are
 taken into account. Extension to include multi-subband effects is straightforward. Each particle interacts
with bulk LO-phonons assumed dispersionless. The total Hamiltonian describing the exciton-LO-phonon complexes
is written as: 

\begin{eqnarray}
H = \sum_{p,{\vec k}} E^p(\vec k) c_{{\vec k}}^{p\dag} c_{{\vec k}}^p + 
\omega_{LO}\sum_{\vec Q}b_{\vec Q}^{\dag}b_{\vec Q} +
\sum_{{p,{\vec k}},{\vec Q}} M_{\vec Q}^p
\frac{1}{|Q|}c_{{\vec k}}^{p\dag} c_{{\vec k}-{\vec q}}^p
\left ( b_{\vec Q} + b_{-\vec Q}^{\dag}\right)
+\sum_{\vec k, \vec k', \vec q}
V^{e-h}_0(q) c_{\vec k' + \vec q}^{e \dag} c_{\vec k'}^{e} c_{\vec k - \vec q}^{h \dag} c_{\vec k}^{h}
\end{eqnarray}
where $E^p(\vec k)$ represents the parabolic dispersion relation for $p$-type
particle ($p=$electron or hole), 
$c_{{\vec k}}^{p,\dag}$ $( c_{{\vec k}}^p)$ is the corresponding creation (annihilation) operator 
with two-dimensional wavenumber $\vec k$, $b_{\vec Q}^{\dag}$ $( b_{\vec Q})$ is the creation (annihilation) operator 
for LO-phonons with momentum $|\vec Q| = (q^2 + q_z^2)^{1/2}$ 
($\vec q$ is the phonon wavevector component paralell to the $xy$ plane),
$M_{\vec Q}^p$ is the particle-phonon interaction matrix element\cite{dasarma}, 
 $V^{e-h}_0$ is the bare electron-hole Coulomb matrix element, with the dielectric constant $\epsilon_0$ replaced by
$\epsilon_\infty$.\cite{ssr,mahan} Herein, $\epsilon_{0}$ and $\epsilon_{\infty}$ are the low- and high-frequency 
limits of the dielectric function, respectively. 

\subsection{First order self-energy corrections}

Due to the large electron- and hole-LO phonon couplings, each carrier interacts with its self-induced
polarization in the polar semiconductor, the so called polaron. Therefore,  the carrier energy must be modified
by self-energy corrections given by\cite{mahan,dasarma}

\begin{eqnarray}
\label{selfener}
\Sigma^{p}({\vec k,\Omega})=-\frac{1}{\beta}\sum_{\vec q,i\nu}V(\vec q,i\nu) {\cal G}_p^0(\vec k -\vec q,\Omega-i\nu)=
\\
\nonumber
\frac{\alpha_p\omega^{3/2}_{LO}}{2\pi (2m_p^*)^{1/2}}
\int \frac{d^2 q}{q}F(q)
\left [
\frac{N_0}{\Omega+\omega_{LO}-E^p
(\vec k-\vec q) }+ 
\frac{N_0 + 1}{\Omega-\omega_{LO}
-E^p(\vec k-\vec q) }
\right ]
\end{eqnarray}
where $V(\vec q,i\nu)$ represents the LO-phonon-mediated effective electron-electron interaction,
${\cal G}_p^0$ is the free $p$-particle propagator, 
$N_0=1/ \left ( e^{\beta \omega_{LO}}-1 \right )$ is the phonon 
population,  $\beta=1/k_B T$. The Fermi factors in Eq.~(\ref{selfener}) are set to zero, because many electron
effects are ignored here. $F(x)$ denotes the
structure factor of the infinite QW 
\begin{eqnarray}
F(q)=\int_0^a dz \int_0^a dz' \phi(z) \phi^*(z) e^{-q|z-z'|}  \phi(z') \phi^*(z')
\end{eqnarray}
with $\phi(z)$ the single particle wave function in the QW growth direction and
  $\alpha_p$ is the Fr\"ohlich coupling 
constant for each kind of particle. The first term 
in the square brackets in Eq.~(\ref{selfener}) represents the absorption of phonons, while the second
one is associated with the emission of phonons. 
Notice that absorption processes can occur only at different from zero temperatures. 

\subsection{Exciton-LO-phonon vertex corrections}

One of the most interesting aspects of the problem is that polarons 
do not interact instantaneously. Therefore, the
real electron-hole interaction, in general, is nonlocal in space and time. This nonlocallity effect is very
important in polar semiconductors modifying dramatically the bare interaction. In order to account for the
effective dressed electron-hole Coulomb interaction, vertex and self-energy corrections 
must be taken into account on an equal footing. The vertex corrections are expressed as\cite{ssr,mahan}
\begin{eqnarray}
I^{e-h}_{eff}(\vec q,i\nu_m)=
\sum_{q_z} \frac{2\pi\gamma}{q^2+q_z^2}F(q_z) 
D_{LO}(\vec Q,i\nu_m)
\label{veffcomplex}
\end{eqnarray}
where $\gamma=e^2 \omega_{LO}\left[1/\epsilon_{\infty} - 1/\epsilon_{0}\right]$
and $D_{LO}(\vec Q,i\nu_m )$ represents the LO-phonon propagator. 

 The effective electron-hole non-local interaction within Shindo's approximation\cite{shindo} is
\begin{eqnarray}
V_{eff}^{e-h}(\vec k,\vec k',\Omega)=\frac{1}{\beta^2}\sum_{i\nu_n,i\nu_n'}
\left [ {\cal G}_e^0(\vec k,\Omega-i\nu_n) + {\cal G}_h^0(-\vec k,i\nu_n) \right] 
I^{e-h}_{eff}(\vec k - \vec k',i\nu_n-i\nu_n') 
\left [ {\cal G}_e^0(\vec k',\Omega-i\nu_n') + {\cal G}_h^0(-\vec k',i\nu_n') \right]
\end{eqnarray}

By summing over the Matsubara frequencies an explicit expression for the effective 
electron-hole non-local interaction results as

\begin{eqnarray}
\label{veff}
V_{eff}^{e-h}(\vec k,\vec k',\Omega)=V_0^{e-h}(\vec k - \vec k')
\left [ 1 + 
\frac{N_0\epsilon_{\infty}\omega_{LO}}{2\epsilon_0} 
\left (\frac{1}{\Omega-E^e(\vec k) -E^h(\vec k')+\omega_{LO}}  + (\vec k \leftrightarrow \vec k')\right )
\right.
\\
\nonumber
+
\left.
\frac{(N_0+1)\epsilon_{\infty}\omega_{LO}}{2\epsilon_0} 
\left (\frac{1}{\Omega-E^e(\vec k) -E^h(\vec k')-\omega_{LO}}  + (\vec k \leftrightarrow \vec k')\right )
\right ]
\end{eqnarray}
From this last expression it becomes obvious that at different from zero temperatures
the effective potential shows singularities at $\Omega = E^e(\vec k) + E^h(\vec k')-\omega_{LO}$. These singularities
increase the exciton-LO-phonon scattering close to $E_X$ for small $k$ values, due to the enhancement of the
phonon density of states in the lower and upper band edges. Eq.~(\ref{veff})
is similar to that found for the three dimensional case, where first order
corrections to the single particle self-energies have been included in the effective electron-hole Coulomb interaction
 screened by an electron-hole plasma.\cite{ssr} 
The second and third terms in the square bracket, Eq.~(\ref{veff}), represent the modification 
of the bare Coulomb potential by phonon effects. These terms describe the dynamical screening of the 
electron-hole interaction and take into account the interband scattering of
electron-hole pairs by absorption or emission of phonons.
Some previous theoretical works\cite{varmasak,koinov,shindo} assumed the static single pole exciton
approximation by replacing $\Omega \rightarrow \Omega-E_X$ both in the carrier self-energy and effective potential. 
Therefore, peaks at $E_X \pm \omega_{LO}$ in the absorption spectrum are expected.
 Within the static single pole exciton approximation and Fourier transforming Eq.(~\ref{veff}) to real 
space a modified Coulomb electron-hole interaction is found, the so-called Haken's potential.\cite{haken} Therefore,
the full expression for the effective potential in Eq.(~\ref{veff}) goes beyond the usual static and
zero temperature effective potential. 

\subsection{Exciton-LO-phonon GF}

Since the creation of electron-hole pairs by the absorption of photons is more efficient in semiconductors of
direct gap, only electron-hole pair states with zero centre-of-mass momentum are considered. 
A systematic way of deriving the exciton-phonon complex properties starts
by considering  the two-particle correlation function.\cite{varmasak,koinov,ssr,shindo} After a long but 
straightforward algebra a Bethe-Salpeter equation for the interacting electron-hole-LO phonon GF is obtained:

\begin{eqnarray}
{\cal G}(\vec k , \vec k',i\nu_n,\Omega) = 
{\cal G}^0(\vec k,\vec k',i\nu_n, \Omega) + 
\frac{1}{\beta}\sum_{\vec k'', \vec k''', i\nu_n'}
 {\cal G}^0(\vec k, \vec k'',i\nu_n,\Omega)
I^{e-h}_{eff}(\vec k'' - \vec k''', i\nu_n-i\nu_n') 
{\cal G}(\vec k''',\vec k',i\nu_n', \Omega)
\end{eqnarray}
with
\begin{eqnarray}
{\cal G}^0 (\vec k, \vec k',i\nu_n,\Omega) = 
\frac {\delta_{\vec k, \vec k'}}
{\left( i\nu_n-E^e(\vec k) - \Sigma^{e}({\vec k,\Omega-i\nu_n})\right ) 
\left ( i\nu_n-E^h(\vec k) - \Sigma^{h}({-\vec k,i\nu_n})\right)}
\label{bethesalpeter}
\end{eqnarray}

The first parenthesis term in the denominator represents the inverse of the dressed by phonons electron propagator 
and the second term the analogous one for a hole.  After summing 
over $\vec k'$, the following
 closed dressed two-particle GF is obtained

\begin{eqnarray}
G(\vec k,i\nu_n,\Omega) = 
G^0(\vec k,i\nu_n, \Omega) + 
\frac{1}{\beta} G^0(\vec k,i\nu_n,\Omega)
\sum_{\vec k', i\nu_n'}
I_{eff}^{e-h}(\vec k - \vec k',i\nu_n-i\nu_n') 
G(\vec k',i\nu_n', \Omega)
\label{grencomplex}
\end{eqnarray}

with
\begin{eqnarray}
G(\vec k,i\nu_n,\Omega) = \sum_{\vec k'} {\cal G}(\vec k , \vec k',i\nu_n,\Omega)
\end{eqnarray}

Using the vertex corrections Eq.~(\ref{veffcomplex}) in the Bethe-Salpeter Eq.~(\ref{grencomplex})
 and performing the frequency summation, the exciton-LO-phonon GF is

\begin{eqnarray}
G(\vec k,\Omega)=G^0(\vec k,\Omega)-G^0(\vec k,\Omega)\sum_{\vec k'}V_{eff}^{e-h}(\vec k,\vec k',\Omega)G(\vec k',\Omega)
\label{gren}
\end{eqnarray}

with

\begin{eqnarray}
G^0(\vec k,\Omega)=\sum_{\vec k',i\nu_n} {\cal G}^0(\vec k,\vec k',i\nu_n,\Omega)=
\frac{1}{\Omega-E^e(\vec k) - E^h(-\vec k)-\Delta_{e-h}(\vec k,\Omega)};\hspace{1cm}
G(\vec k,\Omega)=\sum_{i\nu_n} G(\vec k,i\nu_n, \Omega)
\end{eqnarray}

where 

\begin{eqnarray}
\Delta_{e-h}(\vec k,\Omega)=-\frac{1}{\beta}\sum_{i\nu_n} 
\left[ {\cal G}^0_e(\vec k,\Omega-i\nu_n) + {\cal G}^0_h(-\vec k,i\nu_n) 
\right]
\left[\Sigma^e(\vec k,\Omega-i\nu_n) + \Sigma^h(-\vec k,i\nu_n)\right]
\label{delta}
\end{eqnarray}

The diagrammatic representations of Eqs.~(\ref{selfener}),~(\ref{veffcomplex}) and ~(\ref{grencomplex})
are shown in Fig. 1. 
In order to treat vertex and self-energy corrections on an equal footing, at any temperature,
the following relation (Ward identity) must be held:
\begin{eqnarray}
\Delta_{e-h}(\vec k,\Omega)=\sum_{\vec k'}
\left (V_{eff}^{e-h}(\vec k,\vec k',\Omega) - V_0^{e-h}(\vec k - \vec k')\right )
\label{vertex}
\end{eqnarray}
This last relation is easily probed by using Eqs.~(\ref{selfener}), ~(\ref{veff}), and ~(\ref{delta}). 
\noindent

 Broadening effects, coming from other sources different from LO-phonon
scattering are included phenomenologically by taking $\Omega = \omega + i \delta$. 
The coupled Eqs.~(\ref{selfener}),~(\ref{veff}),~(\ref{gren}), describe the {\it whole exciton-LO-phonon complexes spectrum 
with both discrete and continuum exciton states taken into account}. Temperature effects are considered, both
in the self energy and vertex corrections.
 By solving numerically these equations, the linear absorption spectrum, given by the imaginary part
of the exciton GF  Eq.~(\ref{gren}) as a function of QW width and temperature can be obtained.

\section{Absorption spectrum}

In order to illustrate the behaviour of the exciton-LO-phonon complexes in both the resonant and non-resonant
 regimes in II-VI semiconductor QWs, the absorption spectrum for $ZnSe/ZnSeS$ heterostructures is calculated.
The following parameters are used in the numerical calculations: $m_e=0.16m_0$, $m_h = 0.6 m_0$,
$\epsilon_0=8.8$, $\epsilon_\infty=5.6$, $\omega_{LO}=31.8 meV$.\cite{cingo1} 
 The results should be reliable for well widths $a \leq 100 \AA$, given that only the lowest electron
and highest hole subbands are considered.
Typically the broadening of the
exciton peak ($\delta$) due to impurities, interface roughness, acoustical 
phonons, strain, etc., is roughly the bulk effective Rydberg 
($Ry = 22.4 meV$ for $ZnSe$)\cite{pelekanos2}. In order to
stress the LO-phonon effects the broadening is reduced to $\delta = Ry/2$ 
which was kept fixed for the entire range of temperatures and QW widths. 
Bulk LO-phonons, $i.e.$ unaffected by the well width, are assumed.

\subsection{Continuum resonant state}

First, the exciton absorption in the continuum part of the spectrum is considered. Results for 
different temperatures, $0 K \leq T \leq 330 K$, are discussed. 
In Fig. 2 the absorption spectra, as a function of photon energy for two different 
 well widths, $a=$30$\AA$ (Fig. 2(a)) and $a=$80$\AA$ (Fig. 2(b)) are plotted. Results in Fig. 2(a) correspond
 to a quasi-two-dimensional (Q2D) case while  results appropriate to a quasi-3D case are shown in Fig. 2(b).
The low temperature results provide with an important test of the present formalism:
the exciton effect is quite significant as it should be expected. 
 Besides that, and more important,  an additional peak at $\omega_{LO}$ is present. This feature can be explained
by the fact that the first order carrier self-energy corrections show a logarithmic divergence due to phonon
emission. If vertex corrections were not included the height of this peak would be reduced. 
Additionally, multiphonon contributions, being very weak (they scale like $\approx \alpha^n$)\cite{mahan}, 
 should smooth the resonant state lineshape but do not destroy it.
It is worth noting that the resonant state is still present as a weak shoulder at high enough temperatures. 
However, this peak
 does not correspond exactly to 
the so-called quasi-bound-state at $\Omega=E_X+\omega_{LO}$\cite{toyozawa,hermanson,trallero}. 
In order to recover
this last feature, 
the present formalism could be modified by making the standard
static pole approximation for both the self-energy (Eq.~(\ref{selfener})) and the exciton-phonon
interaction (Eq.~(\ref{veff})), i.e by replacing $\Omega \rightarrow  \Omega-E_X$. 
As a consequence of this modification, the exciton-phonon interaction
develops a strong peak at the exciton binding energy, in agreement with previous theoretical works
\cite{varmasak,koinov,shindo}. 
However, although the present formalism is capable to reproduce well known 
results by using such a phenomenological approach, the main interest in the present
work is related to a systematic perturbative calculation without such kind of procedures.
A full self-consistent treatment must be done in order to place the quasi-bound-state
peak position at the correct energy, i.e. $E_X +\omega_{LO}$. As it is well known, 
such a treatment implies that the self-energy should be calculated in terms of the full
Green's function which in turn depends on the self-energy. 
This kind of calculation has been reported for the three dimensional case in Ref. 14 where a
real space static Haken potential has been used. Nevertheless, dynamic effects in the effective Coulomb
interaction have not been taken into account. By contrast, the present formalism is as a first order 
approximation where the Ward identities are fulfilled and a dynamical effective interaction, for the exciton-phonon
problem, is included.  
However, a full dynamical self-consistent calculation is out of 
the scope of the present work.

\subsection{Exciton peak splitting}

At high enough temperatures, a new result is evident: {\it the exciton peak is splitted}. 
In order to clarify this fact, the inset in Fig. 2b shows the theoretical bare exciton 
binding energy (solid line), $i.e.$ without LO-phonon effects,
and the LO-phonon energy (dotted-line) as a function of the QW width. 
 For a narrow well width, Fig. 2(a), the exciton binding energy is larger than the LO-phonon energy. 
In this case, the number of states into which the exciton can scatter is vanishingly small and consequently 
the effective exciton-LO phonon coupling is reduced. 

Both peaks can be easily identified:  one of them corresponds to the main exciton peak which is red shifted 
as the temperature is increased;  the other one, with a very asymmetrical shape, has an energy position 
 centered at $-\omega_{LO}$ with respect  to the onset of the continuum part of the spectrum. 
Similarly to the continuum resonant state, the asymmetrical lineshape of this new peak should be 
smoothed by including higher order LO-phonon processes. 

Now, in the resonance regime, where a crossing of 
exciton binding-LO-phonon energies occurs, is to be considered.
In Fig. 2b, as the temperature rises from $0 K$ to $330 K$, 
the main exciton peak is Q3D-like and shifts to higher energies. By contrast, the secondary peak preserves
 a constant energy position which is fixed only by the LO-phonon energy. Its intensity becomes
smaller and its linewidth broadens for high temperatures, due to the fact that the density of phonons increases. 
Clearly, these features suggest that the contribution from the phonon absorption terms in the self-energy and
vertex corrections should be observable in the optical absorption spectra.  Similar but weaker peaks have been
 reported by Donlagic et al. \cite{donlagic} for bulk III-V systems using a tight binding model. However, in this 
last case a resonant regime can not be achieved because the binding exciton energy is approximately one half the 
LO-phonon energy.

As the main conclusion of this section, it can be stated that when the temperature is risen, 
the exciton peak splitting should be clearly seen, remaining visible up to $T=330 K$.\cite{pelekanos2} 
In order to see this splitting the Q2D exciton binding energy must be in resonance with the LO-phonon energy
which can be achieved by varying the QW width in II-VI polar heterostructures.

When the crossing of energies occurs ($E_X \approx \omega_{LO}$), the dynamical effective electron-hole pair
interaction becomes very large, and multiple scattering events are dominant (see Eq.~(\ref{gren})). 
From the experimental point of view unexplained shoulders in the exciton main peak absorption have been
reported by Pelekanos et al.\cite{pelekanos} and Miyajima et al.\cite{miyajima}. At the light of the present
theoretical results, those unexplained facts might be associated with non-resolved exciton peak splitting. 
Consequently, to observe experimentally this peak splitting, the samples should be grown in such a way 
that the resonance condition is met  and the absorption spectrum recorded above
the liquid nitrogen temperatures.

\section{Exciton-LO-phonon coupling}

The main points discussed in the previous section 
are: $\it (i)$ an exciton peak splitting and $\it (ii)$ the presence of a continuum resonant state. 
Up to this point, the present model describes succesfully these main facts.
Now the two above mentioned points are to be discussed based on the 
characteristics of the effective electron-hole potential and self-energy effects. 
The fact that the self-energy is included in the free carrier GF implies that 
the effective electron-hole pair interaction dressed by phonons must be treated properly in order to satisfy 
the Ward identity (Eq. ~(\ref{vertex}) with clear consequences in the absorption spectra. In order to
give an ample discussion over the excitonic peak splitting, the effects coming 
 from both self-energy and vertex corrections are to be discussed separately.  

In order to assess the validity of the present results, different contributions from self-energy and
vertex corrections are plotted in Fig. 3. Particularly, the thick solid line, shows the main result of this
work, that is the inclusion of both self-energy and vertex corrections for a 90$\AA$ well width
and $T=200 K$. With these parameters, the excitonic peak splitting is clearly  seen.
 In order to get inside about this splitting, the contribution  from diagrams that violate 
Eq.~(\ref{vertex}) are also considered. 

In the next section, the absorption spectra for
free electron-hole pair recombination, dressed electron-hole pair recombination, 
dressed electron-hole bare Coulomb interaction, electron-hole bare Coulomb interaction, 
electron-hole screened Coulomb interaction
and dressed electron-hole screened Coulomb interaction, are separately displayed in Fig. 3.
Each one of these cases corresponds to a different diagram plotted on the right side of Fig. 3. Notice
that some of them do not satisfy the Ward identity, Eq.~(\ref{vertex}). In spite of that, their study can
contribute to a deeper understanding on the origin of exciton-LO-phonon complexes.

\subsection{Self-energy contributions}

In order to understand  the exciton splitting, the self-energy effects are first 
considered. 
Clearly two peaks at $-\omega_{LO}$ and $\omega_{LO}$ are seen in the absorption spectra, Fig. 3. These peaks
have a strong asymmetric shape due to the singularity of the self-energy. 
The self-energy at $E = E_p(k)\pm w_{LO}$ (Eq.~(\ref{selfener})) has a discontinuity in its
 real part as well as a logarithmic divergence in its imaginary part.
It means that the spectral function shows a strong
singular character. The lineshape of the spectral function might be slightly modified by the
inclusion of second order contributions. 
It is important to highlight that both the resonant carrier-phonon continuum state and the excitonic peak splitting 
come essentially from the dressed single particle properties enhanced by Coulomb interactions. 

For comparison, results with and without self-energy effects are also shown in Fig. 3.
The thin continuous line denotes the case corresponding to the free electron-hole pair recombination, i.e, 
the exciton effects, self-energies and 
vertex corrections are not included. At high energies the absorption spectrum is fully similar to 
a step joint density of states, as it should be. 
The dotted line denotes the case where only self-energy effects are included without Coulomb interaction effects. 
For $T = 200 K$ both absorption and emission of phonons are allowed. Emission effects of the dressed electron-hole pairs
on the absorption spectrum is clearly visible at $\omega_{LO}$ as a shoulder, which will increase 
when Coulomb effecs are added.
However, phonon absorption effects are hardly visible at $-\omega_{LO}$ due to the vanishing optical density 
of the electron-hole pair in the gap region.

Next, the bare Coulomb interaction as well as carrier-phonon dressed effects on the exciton 
absorption (thin dashed line) are taken into account. 
The optical spectrum shows that the exciton peak is red shifted due
to the real part of the self-energy correction $\approx (\alpha_e+\alpha_h)\omega_{LO}/2$ and a new shoulder close to
the exciton peak is now visible. It is important to note
that both the continuum resonance and this new shoulder are enhanced by the Coulomb electron-hole interaction. 

\subsection{Vertex corrections}

In this section the contributions from both the bare Coulomb interaction and vertex corrections to the
absorption spectrum are discussed. In Fig. 3 the thick dashed line represents the typical 
exciton peak with a 3D character ($E_X=-1.43 Ry$). For this case the
binding exciton energy is close to $-\omega_{LO}$.
Now, if the self-energy is not included but the vertex corrections do (dot-dashed line) 
the electron-hole interaction is weakened by the phonon screening and consequently the exciton peak moves
 to the continuum 
part of the spectrum. Shoulders at  $-\omega_{LO}$ and $\omega_{LO}$ disappear, 
confirming that these features come essentially from self-energy contributions. 
Therefore, both self-energy and vertex corrections must be taken into account. 
These corrections are shown as a thick solid-line in Fig. 3. The exciton peak is red shifted,
and the resonances at $E=\pm \omega_{LO}$ are enhanced by the multiple 
electron-hole scattering. 

The temperature dependent effective electron-hole potential $V^{e-h}_{eff}(k,k',\Omega)$,
 averaged over the polar angle between $\vec k$ and $\vec k'$, are plotted in Fig. 4 as a function of the magnitude of one wave vector, $k'$.
Due to the long wavelength of the optical phonons, the effective 
interaction will be more efficient at small $k$. In order to illustrate this effect $k=0.05 a_B$ 
 and a $40 \AA$ well width is considered. Figures 4a, 4b and 4c, show
the effective potential ($Ry$ units) for $\Omega=\omega_{LO}$, $\Omega=0$, $\Omega=-\omega_{LO}$, respectively.
In these figures, the bare effective Coulomb interaction is also plotted for the sake of comparison (thick solid line). 
The bare Coulomb potential is peaked at  $(k=k')$ weighted by the structure factor and a 
similar singularity is found for arbitrary $\Omega$. In Fig. 4a the presence of two
singularities below $k=k'$ can be seen. These singularities, coming from the  first square 
bracket in Eq.~(\ref{veff}), reflect the
fact that at different from zero temperatures the absorption of real phonons by electron-hole pairs is allowed and
must be considered in order to satisfy the Ward identity, Eq.~(\ref{vertex}). 
Decreasing the temperature, these singularities 
dissapear but the third term in Eq.~(\ref{veff}) still remain, enhancing in this way the continuum resonance state 
in the spectra. For 
a static case, Fig. 4b, only one extra singularity is found. 
In Fig. 4c it can be seen that the
effective electron-hole potential is similar to the bare Coulomb interaction
 with $\epsilon_{\infty}$ replacing $\epsilon_0$.  
Clearly, these results reflect the nonlocal character of the effective 
Coulomb interaction and the importance of the vertex corrections at small $k$ values.  Therefore, it is expected that by Fourier
transforming $V^{e-h}_{eff}(k,k',\Omega)$ to the real space the local Haken potential should be modified.\cite{rodriguez}. The degree of this 
nonlocality depends on the energy separation of $\mu*(E\pm\omega_{LO})^{-1}$, with $\mu$ the exciton reduced
mass. 

Experimental results in a related QW system ($ZnCdSe/ZnSe$) present an anomalous plateau in the 
full width at half maximum\cite{cingo1,pellegrini} as a function of the well width. 
 It is tempting to associate this anomalous behaviour with the peak splitting predicted in the resonance region.
However, this last system is in some sense different from the $ZnSe$ based QW calculated here, because in that 
case the QW compound is an alloy and disorder effects must also be taken into account.
This same GF formalism was applied to GaAs QWs, but due to the weak polar character of this last material, 
the main results 
mentioned above could not be seen, confirming the necessity of a highly polar material to observe them. 
 
\subsection{Well width and temperature effects}

Now the optical spectra for different well widths and temperatures are discussed.
By varying the well width the binding exciton energy can be tuned from above the phonon energy to the
resonance region.
 Fig. 5 shows the absorption spectrum at $T=80 K$ (Fig. 5a) and $T=200 K$ (Fig. 5b). 
By increasing the well width, the intensity of 
the excitonic peak decreases,  due to the weakness of the effective electron-hole Coulomb interaction. However,
the splitting of the excitonic peak is still evident. 
For $a = 80\AA$ well width the intensities of both peaks are roughly the same. 
This is a consequence of the fact that vertex corrections and self-energy
are treated on an equal footing, even at finite temperature.
The self-energy and vertex corrections cause a small red shift respect to the bare exciton peak of the
order $\omega_{LO} (\alpha_e+\alpha_h)\pi/2$. 
For $E_X > \omega_{LO}$ it is expected that self-energies corrections are not very important.
This is clear from Fig. 5 for $a = 40 \AA$ where the main contribution comes from
the symmetrical exciton peak, while the secondary peak is weaker and it is located 
 on the high energy side of the main peak. 
By contrast, for a QW width of $a = 80 \AA$, a resonant exciton binding-LO phonon energy crossing
occurs (see inset Fig. 2b), the splitting is evident with the two peak positions exchanged (the secondary
peak on the low energy side of the main peak) but their intensities are still similar.
This splitting is possible because at this temperature there is an important population
of phonons which can be absorbed by the electron-hole pair, which explains why this effect has not
been observed at low temperatures. 
It is worth noting that in both cases and for any well width the continuum resonance is always seen 
as a consequence that LO-phonon emission is allowed as it has been previously reported.\cite{pelekanos}

It is important to stress that this new resonance is different from the well known phonon replica which corresponds
to a peak at $- \omega_{LO}$ below the main exciton peak and
it should be only observable at very low temperature. By contrast,  
the peak splitting predicted here corresponds to a main exciton (symmetric) peak
and a secondary peak shifted by  $- \omega_{LO}$ with respect to the onset of the continuum part of
the spectrum and it should be visible only at high enough temperature. This peak splitting 
could be observed experimentally in the absorption spectrum 
of a similar II-VI QW ($CdTe$) as a shoulder at room temperature in the resonance region, 
but no comment was given about it (see Fig. 3 in Ref.~\ref{pelekanosr}, Fig. 1 in Ref.~\ref{miyajimar}). 
The present results suggest that a splitting of the excitonic
peak can be observed if a systematic control of the inhomogeneous broadening is performed. 
This new peak should be important to understand the 
relaxation and thermodynamic properties of excitons.\cite{umlauff}

On the other hand, a similar peak splitting has been theoretically predicted by 
Castella et al.\cite{castella} invoking disorder
effects in the low temperature regime. The effective electron-hole interaction for 
that case corresponds to a static potential yielding
to elastic exciton scattering events. By contrast, in the present work, a fully non-local and dynamical interaction
is the responsible for the exciton splitting. Moreover, inelastic exciton scattering effects have been taking into
account. 

The formalism used in this work can be extended to study the contribution arising from acoustic 
phonons, both confined and interface-LO-phonons and finite potential QW effects.
 Confined phonon effects in II-VI materials should not be very important since the dielectric
constants of them are very close to each other and therefore bulk LO-phonon energies are very
similar. Hence, the asumption of the bulk LO-phonon energy unaffected by QW width adopted here
should be a reasonable approximation.

\section{Summary}

In summary,  the linear optical exciton absorption of quasi-two
dimensional II-VI systems in the presence of LO-phonons has been studied.
 A Green's function formalism has been developed which takes into account the 
contribution of both discrete and continuum exciton states, at difference
from other works based on the variational formalism. 
For such systems, it should be emphasized that self-energy effects and vertex corrections must be consistently
included on the same footing. 

The main result of this work is the prediction of the 
splitting of the excitonic peak when the following conditions
are met: $(i)$ the exciton binding energy is comparable to the LO-phonon energy and $(ii)$ the temperature
is different from zero. 
The observation of a secondary peak, arising from phonon absorption processes indicates a strong 
exciton-LO-phonon interaction in II-VI QWs. 
This splitting is vaguely visible in III-V systems, because the exciton binding
energy in QWs is always smaller than the LO-phonon energy and the particle-phonon interaction
is weaker than in II-VI systems\cite{donlagic}. 

Besides that striking feature, a shoulder in the continuum part of the optical
spectrum is obtained within the present theoretical approach. 
The calculated absorption spectra are in good agreement with both experimental and previous 
theoretical results. Nevertheless, the secondary peak in the present work is sharper and more 
asymmetrical as compared with the experimental data.
The asymmetrical shape of the secondary peak suggests that LO-phonon processes, 
beyond first-order contributions, and exciton impurity scattering would 
partially modify the lineshape of this new peak. 
Unexplained experimental shoulders in previous works on the exciton main peak absorption could be associated 
with poorly resolved exciton features. 
Although the present procedure has been developed to treat exciton-LO-phonon complexes, it
can also be used to study other effects in low dimensional semiconductors. In particular,
disorder induced effects on the exciton absorption spectrum could be systematically studied.
These studies could lead to a deeper understanding of the linear optical properties of low
dimensional polar systems.

\section{Acknowledgements}

The  author acknowledges  Professor G.D.Mahan and Professor L. Quiroga for the critical reading 
of the manuscript and the stimulating discussions about
the topic of this work. The kind hospitality of the University of Tennessee, where part of this work was
done, and partial finantial support
from the Colombian Institute for Science and Technology (COLCIENCIAS) Project 1204-05-10326 are also acknowledged.
The author also thanks the anonymous referee for his/her comments.

*{frodrigu@uniandes.edu.co}

\newpage

{\bf\Large Figure Captions}

{\bf Fig.1}: 
(a) Dressed by phonons carrier propagator, Eq.~(\ref{selfener}). 
(b) Diagrammatic contribution to the 
effective electron-hole interactions dressed by LO phonons, Eq.~(\ref{veffcomplex}).
(c) Green's function equation for the exciton-LO phonon Eq.~(\ref{grencomplex}).

{\bf Fig.2}: 
Absorption spectrum for $ZnSe$-QW of $a = 30 \AA$ (a)
and $a = 80 \AA$ (b) for different temperatures. 
The inset shows the bare exciton binding energy (solid line) and the bulk LO-phonon energy
(dotted line) as a function
of the well width.
The arrow in the continuum part shows the exciton-phonon resonance of the spectrum.
$E_g$, $E_0^e$ and $E_0^h$ are the band gap, the zero-point energy for electrons and 
holes, respectively. 

{\bf Fig.3}: Absorption spectrum at $T=200 K$ for a $ZnSe$-QW of $a = 90 \AA$. Thin solid line, represents
the free electron-hole pair absorption. Dotted line, is the result of dressing only the free carrier propagators
by LO-bulk phonons. Thick dashed line, represents the exciton absorption without including the vertex and
self energy corrections. Dashed line, displays the excitonic absorption including only the self energy
corrections for free carriers. The dotted-dashed line, represents the excitonic absorption without self-energy
carrier effects but including the LO-phonon vertex correction. The thick-solid line depicts the absorption
spectrum including both self-energy and vertex corrections.

{\bf Fig.4}: Effective dynamical-nonlocal electron-hole Coulomb potential as a function of $k'$ 
for different temperatures at (a) $\Omega=\omega_{LO}$, (b)$\Omega=0$, (c)$\Omega=-\omega_{LO}$ for $k=0.05 a_B$
and well width of $a = 40 \AA$ 


{\bf Fig.5}: Absorption spectrum of $ZnSe$-QW at $T=80 K$ (a) and $T=200 K$ (b) 
 for different well widths. 
The arrow in the continuum part shows the exciton-phonon resonance of the spectrum.
$E_g$, $E_0^e$ and $E_0^h$ are the band gap, the zero-point energy for electrons and 
holes, respectively.


\begin{thebibliography}{9}


\bibitem{ding}
J. Ding et al., Phys. Rev. Lett. {\bf 69}, 11, 5171 (1992).
\bibitem{cingo1}
R.Cingolani et al., Phys.Rev. {\bf B51}, 5176 (1995).
\bibitem{pellegrini}
V.Pellegrini et al., Phys.Rev. {\bf B51}, 5171 (1995).
\bibitem{ehrenreich}
M. E. Flatte et al., Appl. Phys. Lett. {\bf 66}, 1313 (1995).
\bibitem{poweleit}
C. D. Poweleit and L.M. Smith, Phys. Rev. {\bf B55}, 8, 5062 (1997).
\bibitem{varmasak}
S.D. Mahanti and C.M. Varma, Phys. Rev. {\bf B6}, 2209 (1972); 
J. Sak, Phys. Rev. {\bf B6}, 2226 (1972)
\bibitem{pelekanos}
N.T. Pelekanos et al., Appl. Phys. Lett. {\bf 61}, 3154 (1992).
\label{pelekanosr}
\bibitem{pelekanos2}
N.T. Pelekanos et al.,  Phys. Rev. {\bf B45}, 6037 (1992).
\bibitem{oneil}
M. O'Neill et al., Phys.Rev. {\bf B48}, 8980 (1993).
\bibitem{koinov}
Z. Koinov. J. Phys. Cond. Matt. {\bf 3}, 33, 6313 (1991).
\bibitem{pelekanos1}
N.T. Pelekanos et al.,  Phys. Rev. {\bf B56}, R10056 (1997).
\bibitem{toyozawa}
Y. Toyozawa and J. Hermanson,  Phys. Rev. Lett. {\bf B21}, 1637 (1968).
\bibitem{hermanson}
J. Hermanson,  Phys. Rev. {\bf B2}, 5043 (1970).
\bibitem{trallero}
R. Zimmermann et al.,  Phys. Rev. {\bf B56}, 9488 (1997).
\bibitem{umlauff}
M. Umlauff et al., Phys. Rev. {\bf B57}, 1390 (1998).
\bibitem{donlagic}
N. Donlagic and T. \H{O}streich,  Phys. Rev. {\bf B59}, 7493 (1999).
\bibitem{ssr}
H. Haug and S. Schmit-Rink, Prog. Quant. Electr. {\bf 9}, 3 (1984).
\bibitem{mahan}
G.D. Mahan, {\sl Many Particle Physics}, (Plenum Press, New York, 1990).
\bibitem{castella}
H. Castella and J. Wilkins, Phys.Rev. {\bf B58}, 16186 (1998).
\bibitem{zheng}
R. Zheng and M. Matsuura, Phys.Rev. {\bf B58}, 10769 (1998)
\bibitem{gerlach}
B.Gerlach et al., Phys.Rev. {\bf B58}, 10568 (1998)
\bibitem{chuu}
D. S. Chu et al., Phys.Rev. {\bf B49}, 14554 (1994)
\bibitem{frolich}
H. Fr\"ohlich, Advances in Physics {\bf 3}, 325 (1954).
\bibitem{dasarma}
S. Das Sarma et al., Annals of Physics, {\bf 163}, 78 (1985)
\bibitem{shindo}
K.Shindo, J. Phys. Soc. Japan, {\bf 29}, 287 (1970).
\bibitem{haken}
H. Haken, Fortschr. Phys. {\bf 6}, 271 (1958).
\bibitem{miyajima}
Miyajima et al., Appl. Phys. Lett. {\bf 66}, 2 (1995).
\label{miyajimar}
\bibitem{rodriguez}
F. J. Rodr\'{\i}guez, to be published.
\label{rodriguez}

\end{thebibliography}
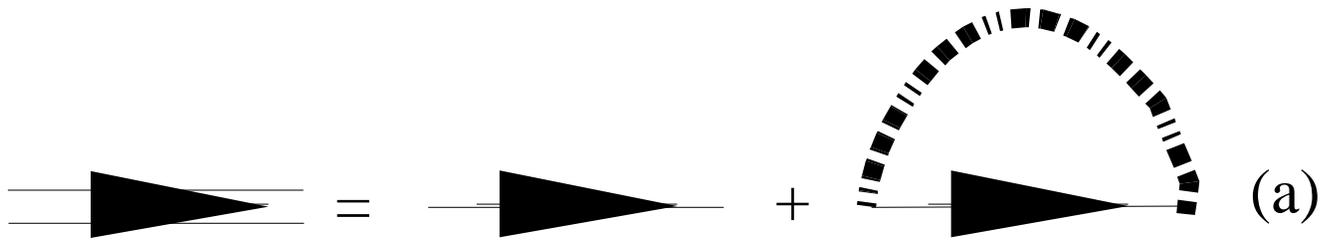
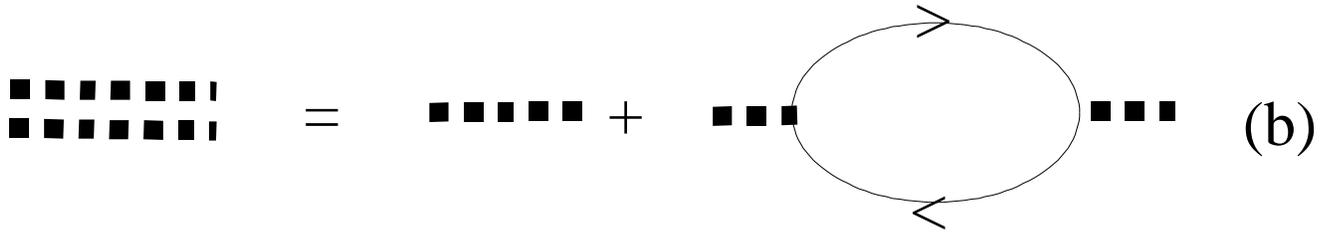
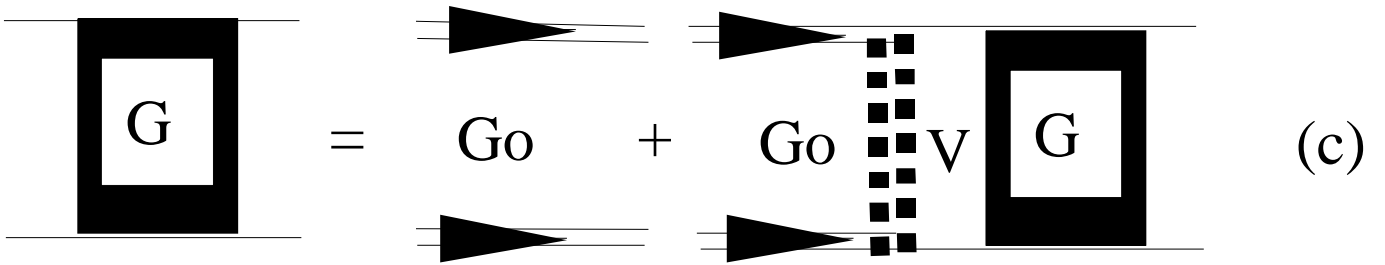
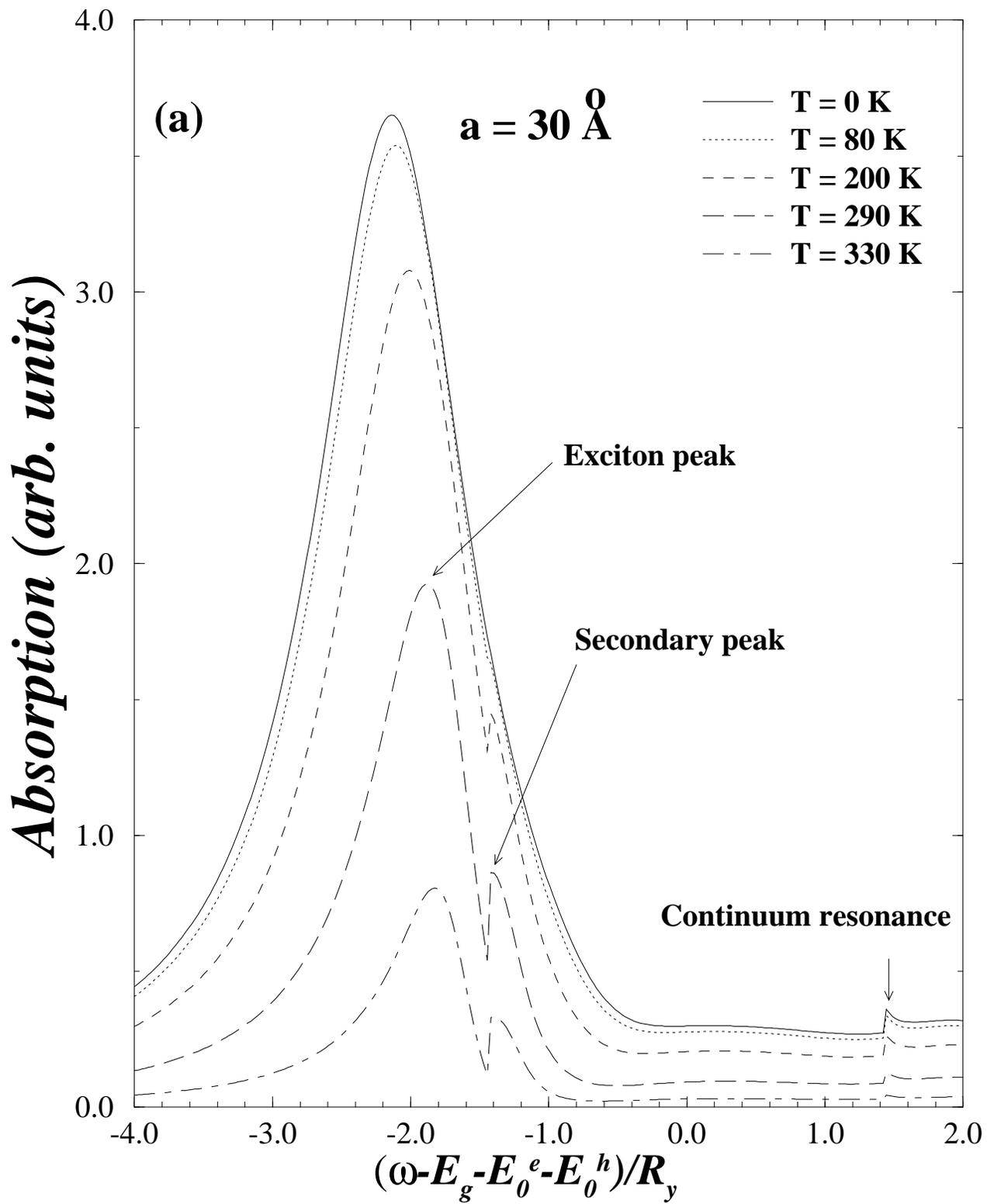
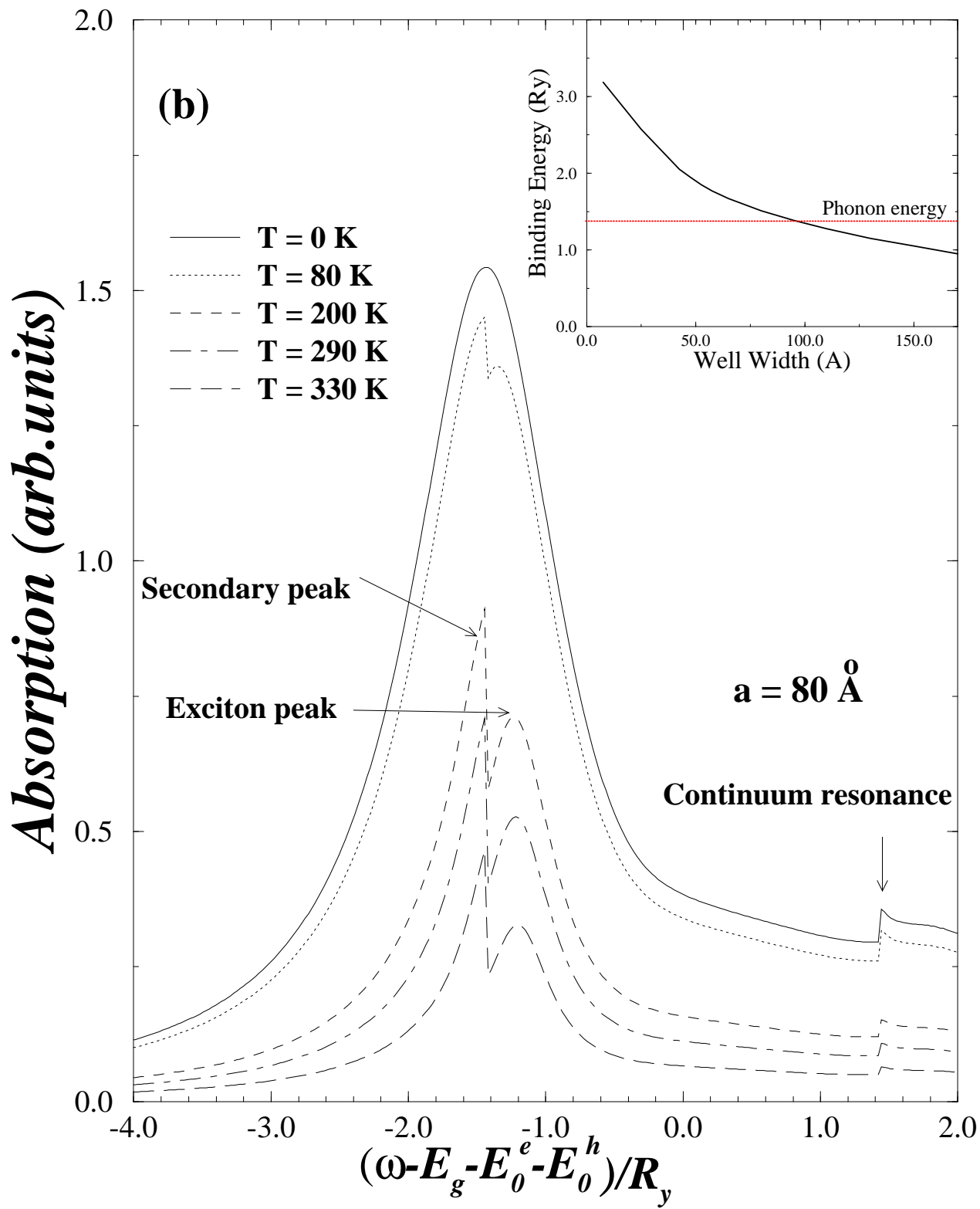
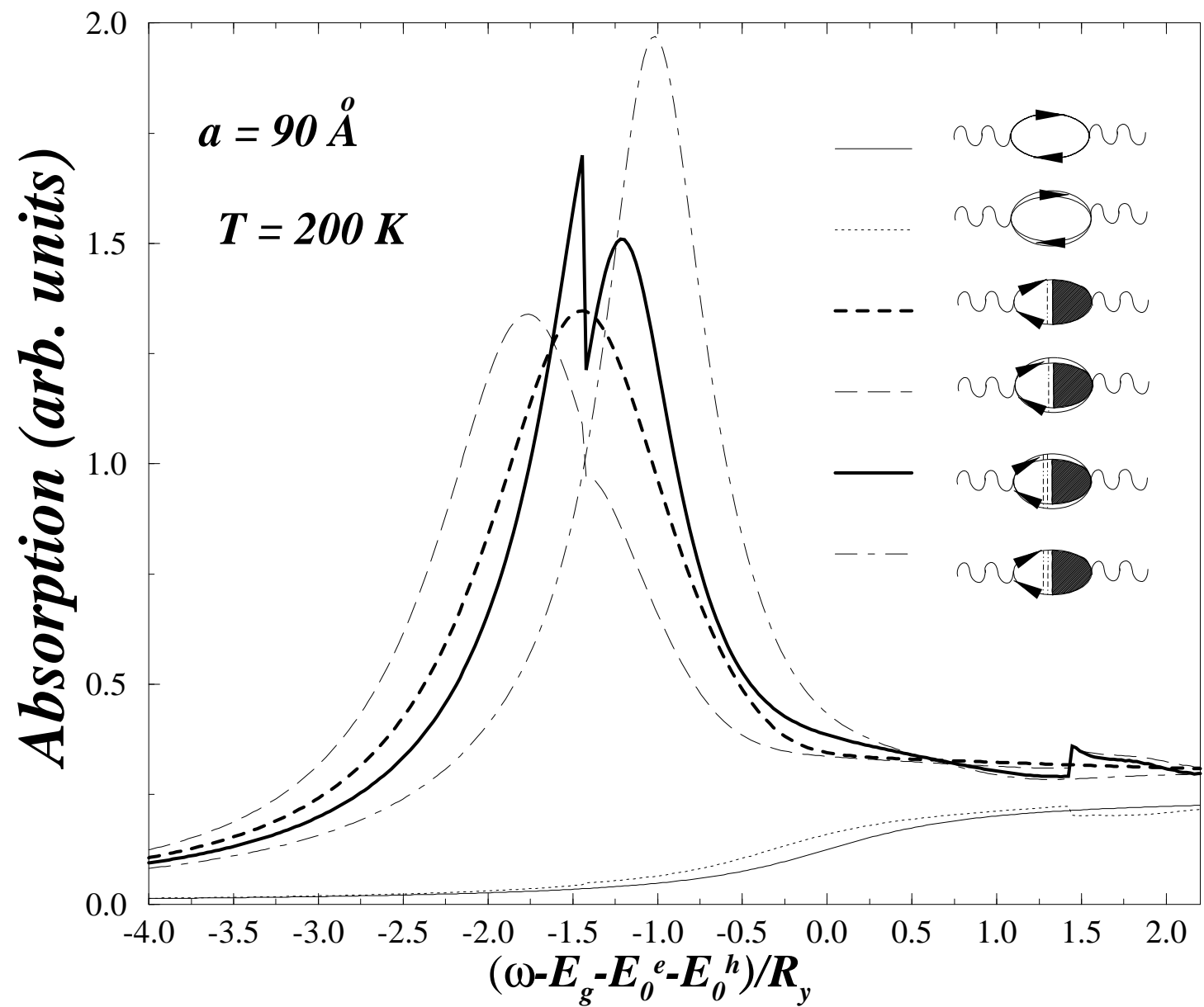
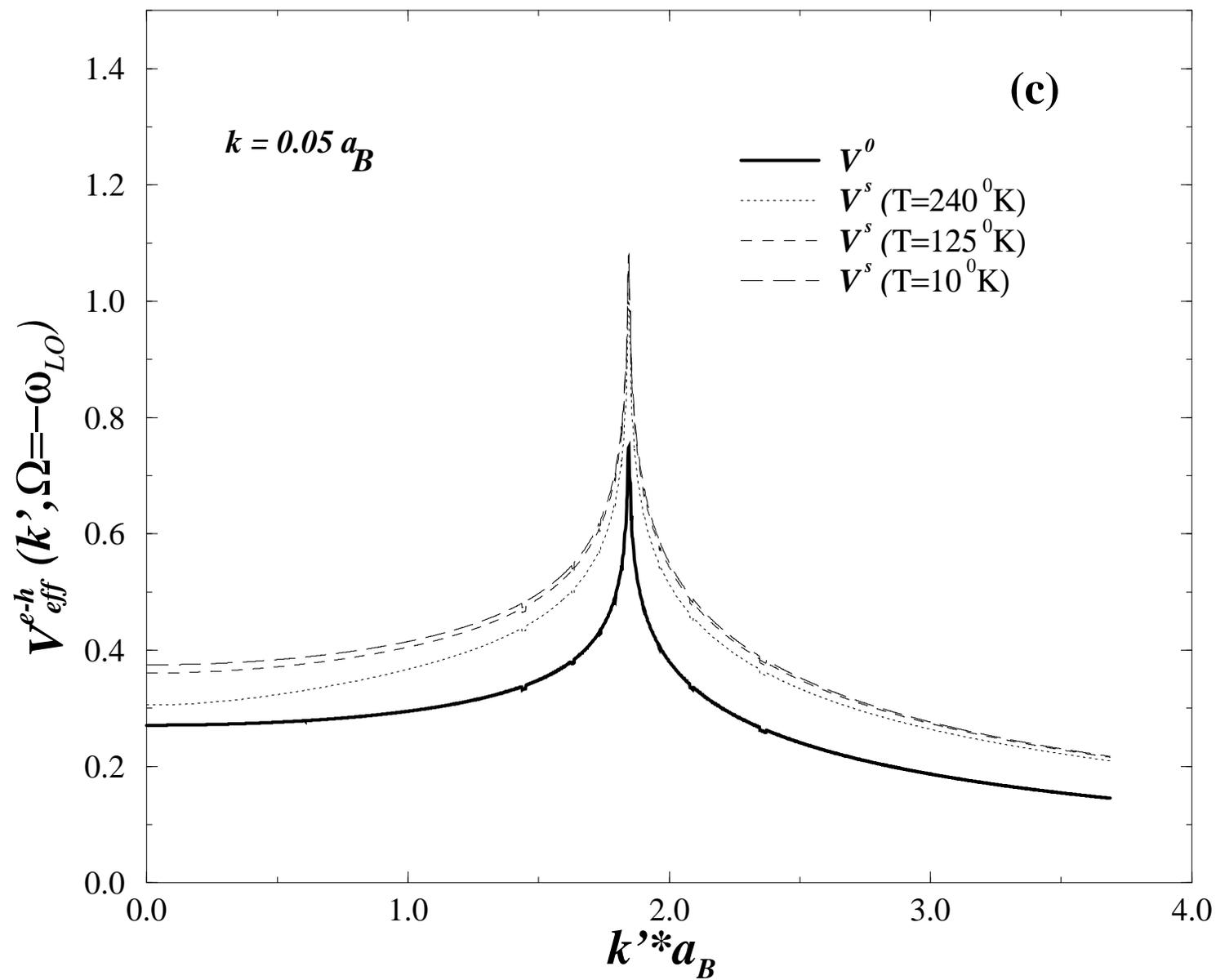
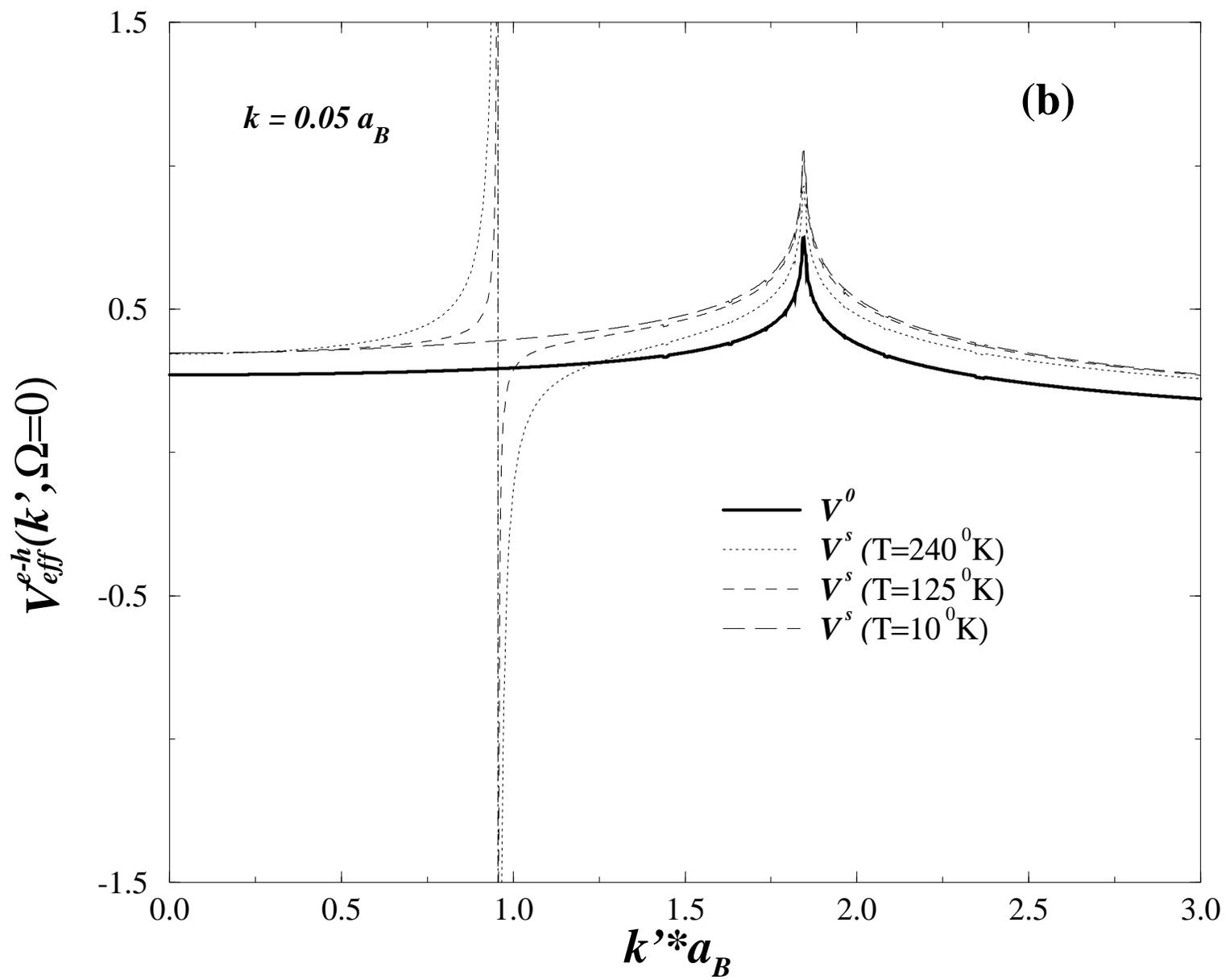
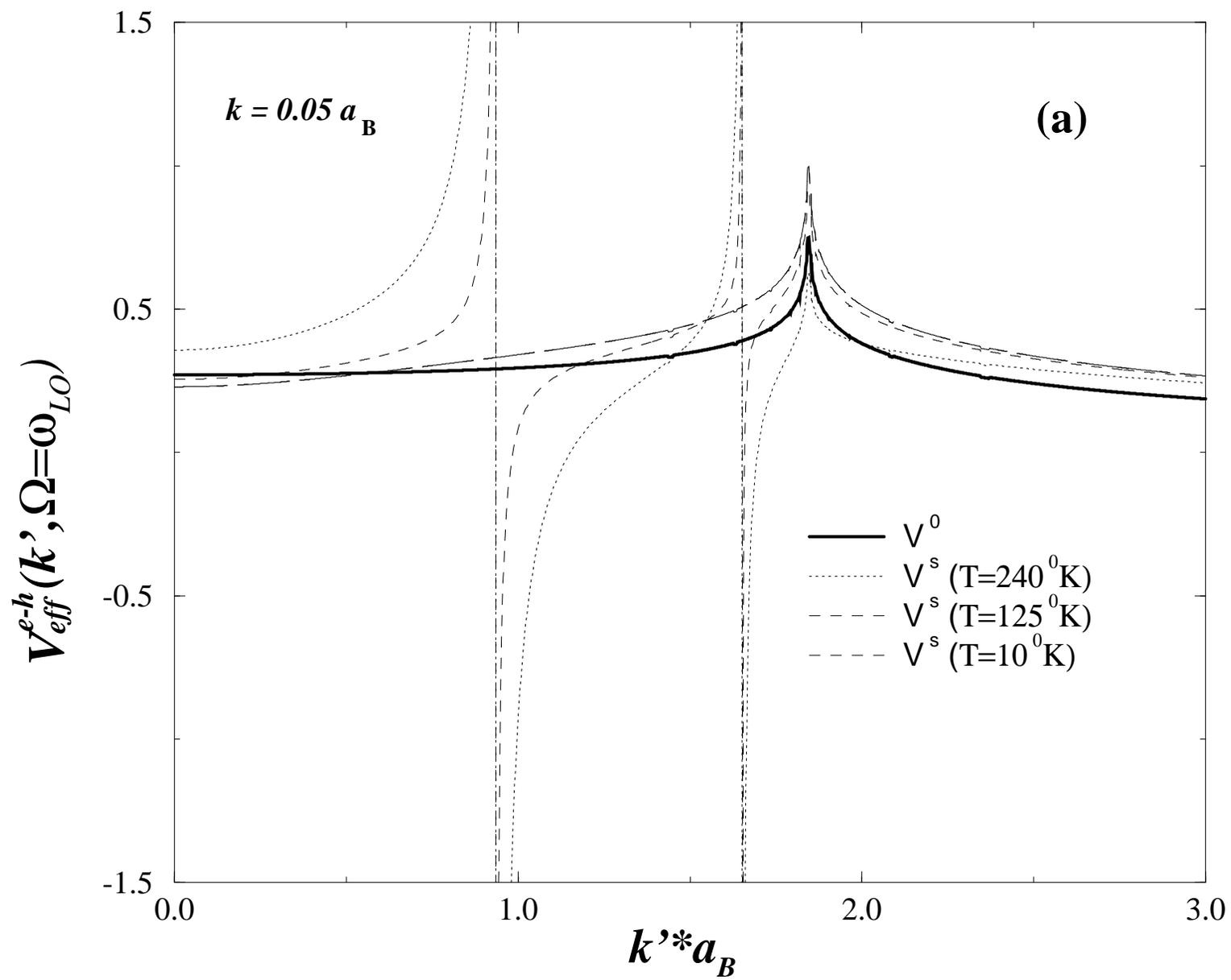
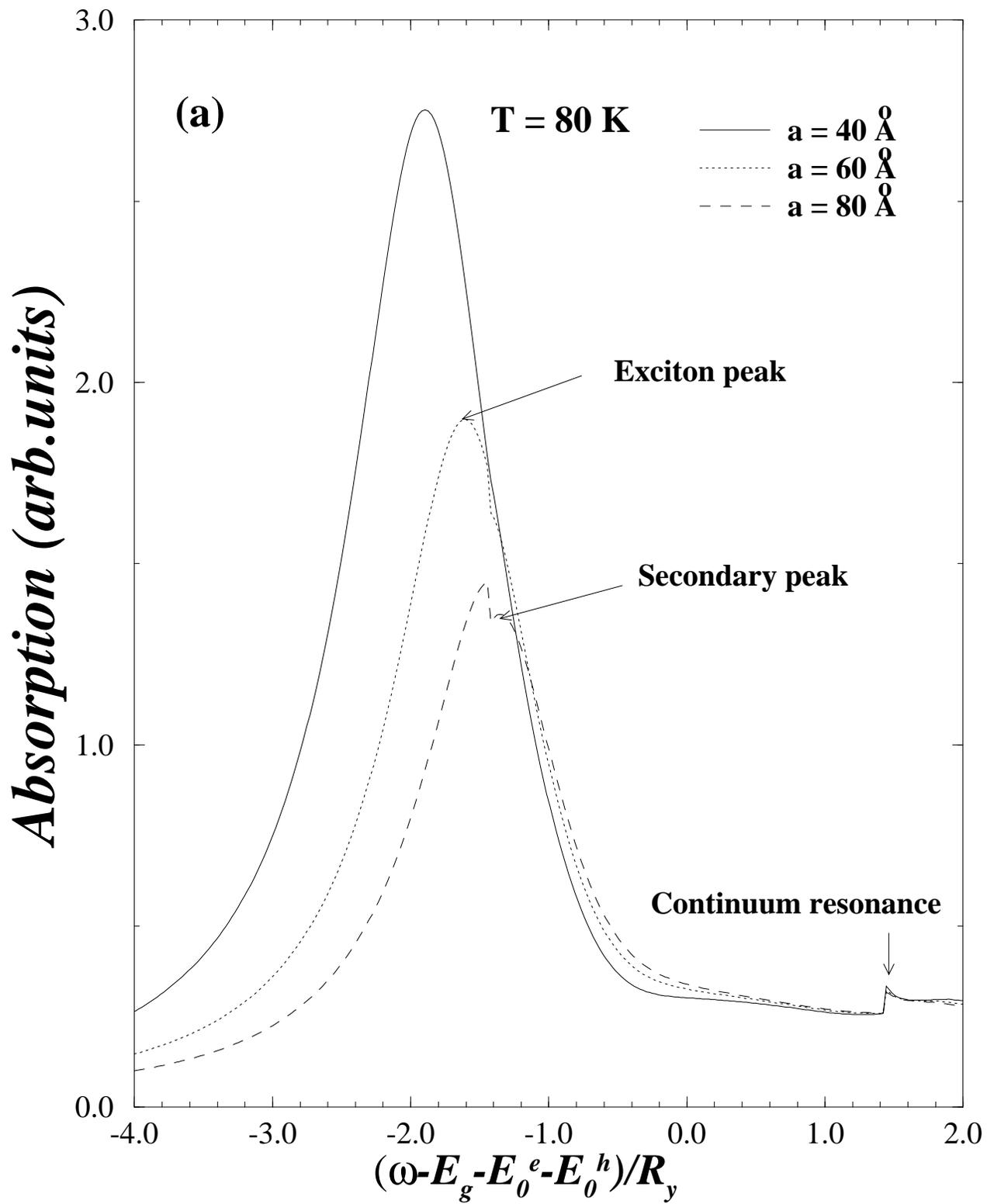
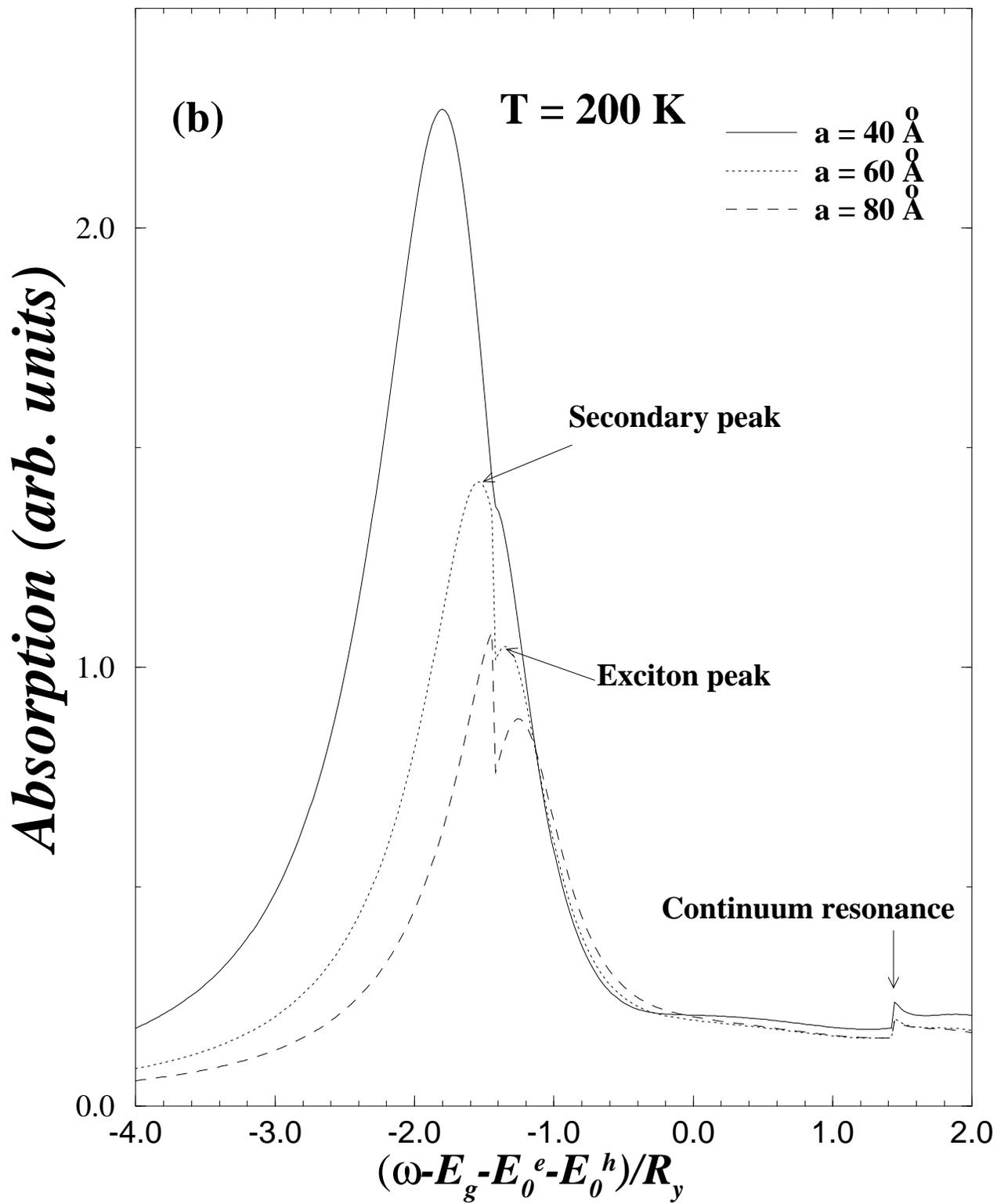
\end{document}